\documentclass[prd,twocolumn,preprintnumbers,superscriptaddress,nofootinbib]{revtex4}
\usepackage[a4paper, hdivide={1.91cm,,1.165cm}, vdivide={1.83cm,,2.6cm}]{geometry}

\usepackage{amstext,amssymb}
\usepackage{amsmath}
\usepackage{graphicx}
\usepackage[hyperfootnotes=false]{hyperref}
\usepackage{xspace}
\usepackage{color}
\usepackage{units}
\usepackage{slashed} 
\newtheorem{theorem}{Theorem}

\begin{document}

\title{\hspace*{-1.0cm} Theorem on vanishing contributions to $\sin^2\theta_W$ and intermediate mass scale in Grand Unified Theories with trinification symmetry}


\author{Chandini  \surname{Dash}}
\email{dash25chandini@gmail.com}
\affiliation{Department of Physics, Berhampur University, Odisha-760007, India.}


\author{Snigdha  \surname{Mishra}}
\email{mishrasnigdha60@gmail.com}
\affiliation{Department of Physics, Berhampur University, Odisha-760007, India.}

\author{Sudhanwa \surname{Patra}}
\email{sudhanwa@iitbhilai.ac.in}
\affiliation{Department of Physics, Indian Institute of Technology Bhilai, India}

\begin{abstract}

We prove that the values of the electroweak mixing angle $\sin^2\theta_W$ and intermediate mass scale $M_I$ have vanishing contributions due to one-loop, two-loop and gravitational corrections in Grand Unified Theories which accommodate an intermediate trinification symmetry ($G_{333D}$) invoked with spontaneous D-parity mechanism operative at mass scale greater than  $M_I$. The proof of theorem is robust and we verify the results numerically using supersymmetric as well as non-supersymmetric version of $E_6$-GUT.
\vspace*{-1.0cm}
\end{abstract}


\maketitle

\section{Introduction}
\noindent
The profound discovery of Higgs boson at Large Hadron Collider glorifies the success of the Standard Model (SM) and the importance of spontaneous symmetry breaking. The final step of spontaneous symmetry breaking for all Grand Unified Theories (GUTs) have to proceed through the SM gauge group and to reproduce all SM observables. The theoretical and experimental determination of known SM observables including the electroweak mixing angle $\sin^2\theta_W$ is more meaningful if the theoretical predictions from GUT models are consistent with these experimentally measured values. Towards achieving this goal, many theoretical GUTs models--like $SU(5)$\cite{Georgi:1974sy}, $SO(10)$\cite{Pati:1974yy,Fritzsch:1974nn,Mohapatra:1974gc,Mohapatra:1974hk,Senjanovic:1975rk,DeRujula:1977wll,Marshak:1979fm,Clark:1982ai,Aulakh:1982sw}, $E_6$\cite{Gursey:1975ki,Gursey:1976dn,Sikivie:1977st,Nandi:1985uh,Stech:2003sb} etc with or without superysmmetry (SUSY), have been proposed as prospective theories of nature awaiting experimental feasibility. With the development of String Theories \cite{Ma:1987zm,Gunion:1987xi,Hewett:1988xc} the attention is now more pointing towards re-investigating the Exceptional Group $E_6$ as a more plausible choice, as it accommodates most of the nice features of the well known GUT groups $SU(5)$, $SU(6)$ and $SO(10)$ through it's embedding. The $E_6$ GUT, started in the late 1970's(just after $SO(10)$ GUT models) is primarily introduced to explain the hierarchy problem, the mysterious anomaly cancellations of the Standard Model and the quantization of electric charge etc. Many of the pressing issues of SM can be adequately addressed here, with key predictions on achieving gauge unification, stability of proton, strong CP problem, predictions of dark matter candidature, non-zero neutrino masses and mixing etc. Some of the above phenomenological aspects have been discussed for $E_6$ gauge model via its maximal subgroups $SO(10)\otimes U(1)$\cite{Howl:2007zi} and $SU(3)_C\otimes SU(3)_L\otimes SU(3)_R$\cite{Babu:1985gi,Mohapatra:1985xm,Nishimura:1988fp,Willenbrock:2003ca,Sayre:2006ma,Cauet:2010ng,Stech:2014tla,Pelaggi:2015kna}.   

  The usual GUT predictions on these observables using standard formalism can be derived by solving renormalization group equations (RGEs) by taking into account one-loop~\cite{Gross:1973id}, two-loop effects~\cite{Jones:1974mm,Jones:1981we} as well as gravitational ~\cite{Shafi:1983gz,Hill:1983xh,Parida:1989pq,Patra:1991dy} and GUT-threshold corrections\cite{Parida:1990wm,Mohapatra:1992dx,Parida:2002rv}. The gravitational and threshold corrections can arise due to presence of higher dimensional operators or due to presence of heavy particles contained in large representations sitting at/around GUT scale. Here it is noteworthy to mention that the stability of the intermediate mass scale $M_{\rm I}$ and $\sin^2\theta_W$ is essential for neutrino mass predictions through seesaw mechanism and other relevant phenomenological observables. In this context, we aim to demonstrate that there is cancellation of all corrections due to one-loop, two-loop and gravitational effects for the intermediate mass scale $M_{\rm I}$ and the electroweak mixing angle $\sin^2\theta_W$ in a class of GUT models accomodating intermediate trinification symmetry 
$SU(3)_C\otimes SU(3)_L\otimes SU(3)_R$ invoked with D-parity $(G_{333D})$ . It has been shown earlier the vanishing contributions~\cite{Parida:1990wm,Parida:1992sj,Mohapatra:1992jw,Parida:1996td} due to one-loop, two-loop, gravitational as well as threshold in case of intermediate Pati-Salam symmetry~\cite{Pati:1974yy,Mohapatra:1974gc,Senjanovic:1975rk} with D-parity\cite{Chang:1983fu,Chang:1984uy}. However, all these corrections survive for the unification mass scale $M_{U}$ as well as GUT coupling constant.\\

This paper is organised as follows. In Section-II, we propose a general theorem alongwith its proof for vanishing corrections due to one-loop, two-loop and gravitational effect for the intermediate mass scale $M_{\rm I}$ and the electroweak mixing angle $\sin^2\theta_W$ in a class of GUT models with intermediate  $G_{333D}$ symmetry. 
We verify the theorem numerically in $E_6$ GUT model with and without suspersymmetry in the consequent Section-III and IV respectively. The last Section is devoted to  the concluding remarks alongwith a comment on threshold corrections.
  

\section{Theorem and Its Proof}
\vspace*{-0.1cm}
\begin{theorem}
\noindent \\
In a class of Grand Unified Theories (GUTs) having trinification symmetry 
$SU(3)_C\otimes SU(3)_L\otimes SU(3)_R\otimes D$ as the highest intermediate symmetry, the electroweak mixing angle $\sin^2\theta_W$ and the  intermediate trinification symmetry breaking scale ($M_{\rm I}$) have vanishing contributions due to one-loop, two-loop and gravitational corrections arising from mass scales greater than the intermediate scale (i.e., $\mu > M_{\rm I}$).     
\end{theorem}
{\em{Proof:-}} \,
In order to prove the theorem, we consider symmetry breaking chain of a GUT model with intermediate trinification symmetry as
\begin{eqnarray}
\vspace*{-0.2cm}
\mathbb{G}_{\rm GUT} &\stackrel{M_U}{\longrightarrow}& SU(3)_C\otimes SU(3)_L\otimes SU(3)_R\otimes D (G_{333D})\nonumber \\ 
	&\stackrel{M_I}{\longrightarrow}&SU(3)_C\otimes SU(2)_L\otimes U(1)_{Y} (G_{321})\nonumber\\
	&\stackrel{M_Z}{\longrightarrow}& SU(3)_C\otimes U(1)_{Q} (G_{31})
	\end{eqnarray}
Here, $D$ stands for D-parity known as discrete left-right symmetry. We mainly use $E_6 \equiv \mathbb{G}_{\rm GUT}$ in proving the theorem numerically, whereas $\mathbb{G}_{\rm GUT}$ can be any GUT model which can accommodate trinification gauge symmetry as highest intermediate symmetry. 

Now to evaluate the loop-effects, we follow the standard procedure by using the renormalization group equations (RGEs)\cite{Georgi:1974yf}. The evolution of gauge coupling constant $g_i (\mu)$ of an intermediate gauge symmetry $\mathbb{G}_i$ occurring in $\mathbb{G}_{\rm GUT} \to G_{\rm I} \to \mathbb{G}_{\rm SM}$ is given by,
\begin{equation}
\mu\,\frac{\partial g_{i}}{\partial \mu}=\frac{\pmb{b_i}}{16 \pi^2} g^{3}_{i}+
\frac{1}{(16 \pi^2)^2}\, \sum_{j} \pmb{B_{ij}}_{} g^3_{i} g^2_{j}\, .
\label{rge-coupl}
\end{equation}
After simplification, the known analytic formula derived for one loop as well as two loop RGEs for inverse coupling constant valid from $\mu$ to the intermediate scale $M$ ($M$ can be of any scale $> \mu$ where new theory appears) as,
\begin{eqnarray}
\frac{1}{\alpha_{i}(\mu)} &=& \frac{1}{\alpha_{i}(M)} + \frac{\pmb{b_i}}{2 \pi}\,  \text{\large ln}\left(\frac{M}{\mu} \right) \nonumber \\
&&\hspace*{0.5cm}+\frac{1}{8 \pi^2}\, \sum_{j} \pmb{B_{ij}} \int^{M}_{\mu} \alpha_j(\mu) \frac{d\mu}{\mu} \, .
\label{rge-alphainv}
\end{eqnarray}
Here, $\alpha_{i}=g^2_{i}/(4 \pi)$, where $g_i$ being the coupling constant for the $i^{\rm th}$ gauge group. $\pmb{b_i}$ ($\pmb{B_{ij}}$) is the one (two)-loop beta coefficients in the mass range $\mu-M$.

The RGEs for evolution of gauge coupling constant in between the mass scale $M_Z$ and $M_I$ are given by
\begin{equation}
\hspace*{-0.4cm}\alpha_i^{-1}(M_Z)=\alpha_i^{-1}(M_I)+\frac{\pmb{b_i}}{2\pi}{\ln\left(\frac{M_I}{M_Z}\right)}+ \pmb{\Theta_{i}}
\label{rge-alphainv-muMI}
\end{equation}
with $i=\mbox{1Y, 2L, 3C}$. However, there is an additional gravitational corrections from $M_I$ and $M_U$ scale emerging from non-renormalizable higher dimensional operators. The corresponding RGEs for evolution of gauge coupling constants are,
\begin{equation}
\hspace*{-0.4cm}\alpha_i^{-1}(M_I)=\alpha_i^{-1}(M_U)+\frac{\pmb{b^\prime_i}}{2\pi}{\ln\left(\frac{M_U}{M_I}\right)}+ \pmb{\Theta^{\prime}_{i}} - \pmb{\Delta^{\rm \small NRO}_i}
\label{rge-alphainv-MIMU}
\end{equation}
where, $i=\mbox{3C, 3L, 3R}$. The second and third term in the RHS of both eqns.  (\ref{rge-alphainv-muMI}) and (\ref{rge-alphainv-MIMU}) represent the one-loop effects and two-loop effects respectively while the fourth term of equation (\ref{rge-alphainv-MIMU}) is for gravitational contributions. Here $\pmb{b_i}$ ($\pmb{b^\prime_i}$) is the one-loop beta coefficients derived for particle spectrum within mass range $M_Z-M_I (M_I-M_U)$. In the above mass range,the two-loop contributions are $\pmb{\Theta_{i}}$ and $\pmb{\Theta_{i^{'}}}$, respectively, as given by
\begin{eqnarray}
\pmb{\Theta_{i}}&=&\frac{1}{4\pi}\sum_j \pmb{B_{ij}} \ln\frac{\alpha_j(M_I)}{\alpha_j(M_Z)}\, , \quad (\mbox{for}\quad M_Z-M_I) \nonumber \\
\pmb{\Theta^\prime_{i}}&=&\frac{1}{4\pi}\sum_j \pmb{B^\prime_{ij}} \ln\frac{\alpha_j(M_U)}{\alpha_j(M_I)} \, , \quad (\mbox{for}\quad M_I-M_U)
\label{two-loop}
\end{eqnarray}
with, 
\begin{eqnarray}
 \pmb{B_{ij}}=\frac{\pmb{b_{ij}}}{\pmb{b_j}}, \pmb{B^\prime_{ij}}=\frac{\pmb{b^\prime_{ij}}}{\pmb{b^\prime_j}}
 \label{two-loop-beta}
\end{eqnarray}
Here, $\pmb{b_{ij}}$ ($\pmb{b^\prime_{ij}}$) is the two-loop beta coefficients derived for particle spectrum within mass range $M_Z-M_I (M_I-M_U)$. The role of $\pmb{\Delta^{\rm \small NRO}_i}$ in eq.(\ref{rge-alphainv-MIMU}) is to modify the GUT scale boundary condition at $\mu=M_U$ as,
\begin{eqnarray}
\alpha_{i} (M_U)(1+\epsilon_{i})=\alpha_{G} \, \quad \mbox{with}~\,  \mbox{i=3C, 3L, 3R}\,
\end{eqnarray}
Here $\epsilon_i$ is the parameter which induces the gravitational correction.\\
 Thus the gravitational contributions $\pmb{\Delta^{\rm \small NRO}_i}$ can be put in simplified form, 
\begin{eqnarray}
 \Delta_i^{NRO}=-\frac{\pmb{\epsilon_i}}{\alpha_G} \, \quad \mbox{with}~\,  \mbox{i=3C, 3L, 3R}\,
\end{eqnarray}
where $\alpha_G$ is the GUT-coupling constant. 
Here it may be noted that the gravitational contribution is due to the non-renormalizable term\cite{Rizzo:1984mk} given by
\begin{eqnarray}  
&&\mathbb{L}_{NRO}=-\frac{\eta}{4 M_G} \textbf{Tr} \Big(F_{\mu\nu} \pmb{\Phi} F^{\mu\nu}\Big)
\label{eq:NRO}
\end{eqnarray}
where $\pmb{\Phi}$ is the Higgs field responsible for breaking of the GUT symmetry $G$ to $G_{333D}$ at the mass scale $M_U$. The gravitational correction parameter $\epsilon_i$ depends on $\eta$, VEV of $\phi$ and $M_G$. Here it is noteworthy to mention that the  non-renormalizable term in the Lagrangian may arise as effects of quantum gravity\cite{Hill:1983xh,Rizzo:1984mk} or as a result of compactification of extra dimensions\cite{Shafi:1983gz}. Thus in eq.(\ref{eq:NRO}), the scale $M_G\simeq M_{Pl}\simeq 10^{19}$ GeV, if it is due to quantum gravity effect. However, if it emerges as a result of compactification of extra dimensions then $M_G\leq M_{Pl}$ \cite{Shafi:1983gz}.

\noindent \\ \\
\underline{\bf Analytical proof for $\sin^2\theta_W$:-}
Using the standard procedure, the analytic formula for $\sin^2\theta_W$ due to contributions arising from one-loop, two-loop and gravitational correction is given by
\begin{eqnarray}
\sin^2\theta_W&&= \frac{1}{A_U}\Big[\frac{3}{8}A_U +\left(\frac{\alpha_{\rm em}}{\alpha_s}-\frac{3}{8}\right)B_U \nonumber \\
&&+ \frac{\alpha_{\rm em}\left(A_{U}B_{I}-A_{I}B_{U}\right)}{16\pi}\ln\left(\frac{M_I}{M_Z}\right) \Big]
\nonumber \\
&&+ \frac{1}{A_U}\Big[\frac{\alpha_{\rm em}\left(A_{U} K_{\pmb{\Theta} \pmb{\Theta^\prime}} -B_{U} J_{\pmb{\Theta} \pmb{\Theta^\prime}}\right)}{16\pi} \Big] \nonumber \\
&&+\frac{1}{A_U} \Big[\frac{\alpha_{\rm em} \left(A_{U} E_{0} - B_{U} E_{1}\right)}{8 \alpha_G}\Big]
\label{rel:sinsqthetaw}
\end{eqnarray}
The first, second and third squared-bracketed terms in (\ref{rel:sinsqthetaw}) are due to one-loop, two-loop and gravitational effects repspectively. Now in order to prove the theorem, we need to focus on each term separately.
\begin{itemize}
 \item The one-loop contributions to the electroweak mixing angle $\sin^2\theta_W$ is given by
 \begin{eqnarray}
 \hspace*{0.8cm}\left( \sin^2\theta_{W}\right)_{1-loop}&&=  \frac{1}{A_U}\Big[\frac{3}{8}A_U\hspace*{-0.1cm}+\hspace*{-0.1cm}\left(\frac{\alpha_{\rm em}}{\alpha_s}-\frac{3}{8}\right)B_U \nonumber \\
 &&
\hspace*{-0.1cm}+\hspace*{-0.1cm} \frac{\alpha_{\rm em}\left(A_{U}B_{I}-A_{I}B_{U}\right)}{16\pi}\ln\left(\frac{M_I}{M_Z}\right) \Big]\, . \nonumber
\end{eqnarray}
 Here the parameters $A_I$, $B_I$, $A_U$ and $B_U$ $\left(\mbox{as can be inferred from appendix}\right)$ contains the one-loop effect at $\mu > M_{\rm I}$. However in the first term $A_U$ cancelled out in both numerator as well as in denominator which leads to $3/8$ only. Second term completely vanishes as $B_{U}= 4 \pmb{b^\prime_{3L}}-4 \pmb{b^\prime_{3R}}=0$ since $\pmb{b^\prime_{3L}}=\pmb{b^\prime_{3R}}$ in all class of GUT models with intermediate trinification symmetry invoked with D-parity ($G_{333D}$).  With $B_U=0$ the third term contains the parameter $B_{\rm I}=5 \pmb{b_{2L}}-5 \pmb{b_{Y}} \left(\mbox{as has been shown in appendix}\right)$, here we may note that $B_I$ has only one-loop effect at $\mu < M_{\rm I}$. Thus the modified expression for $\sin^2\theta_W$ is given by
\begin{equation}
\hspace*{0.8cm} \left( \sin^2\theta_{W}\right)_{1-loop}= \frac{3}{8}+\frac{\alpha_{\rm em} (5b_{2L}-5b_{Y})}{16\pi}\ln\left(\frac{M_I}{M_Z}\right) \, . \nonumber
\end{equation}
which shows that $\sin^2\theta_W$ has vanishing contributions from one-loop effects emerging form $\mu > M_{\rm I}$.
\item The two-loop contributions to the electroweak mixing angle is given by
\begin{eqnarray}
\hspace*{0.8cm}\left( \sin^2\theta_{W}\right)_{2-loop} = \frac{1}{A_U}\Big[\frac{\alpha_{\rm em}\left(A_{U} K_{\pmb{\Theta} \pmb{\Theta^\prime}} -B_{U} J_{\pmb{\Theta} \pmb{\Theta^\prime}}\right)}{16\pi} \Big] \nonumber
\end{eqnarray}
Now using the similar analysis as in case of one-loop, we can see that the above expression reduces to $\frac{\alpha_{\rm em} K_{\pmb{\Theta} \pmb{\Theta^\prime}} }{16\pi}$, where $K_{\pmb{\Theta} \pmb{\Theta^\prime}}$ has been defined in the eq.(\ref{app:Ktheta}). However $K_{\pmb{\Theta} \pmb{\Theta^\prime}}$ is independent of one-loop contribution emerging from mass scale $\mu > M_{\rm I}$ since $\pmb{\Theta^\prime_{3L}}=\pmb{\Theta^\prime_{3R}}$. This is quite natural in all GUT models with conserved D-parity. Thus $\sin^2\theta_W$ has vanishing contributions from two-loop effects emerging form $\mu > M_{\rm I}$. 
\item The important gravitational contribution to $\sin^2\theta_W$ is as follows,
\begin{equation}
 \left( \sin^2\theta_{W}\right)_{NRO}= \frac{1}{A_U} \Big[\frac{\alpha_{\rm em} \left(A_{U} E_{0} - B_{U} E_{1}\right)}{8 \alpha_G}\Big]\, . \nonumber
\end{equation}
Here we may note that for $B_U=0$, the second term vanishes identically. In the first term $A_U$ cancelled out from both numerator and denominator. Again $E_0\left(= 4\pmb{\epsilon_{3L}}-4\pmb{\epsilon_{3R}}\right)$ vanishes due to D-parity conservation. As a result of this, there is no effect of gravitational corrections at all to the electroweak mixing angle $\sin^2\theta_W$ i.e, $\left( \sin^2\theta_{W}\right)_{NRO}=0$.  
\end{itemize}

\noindent \\
\underline{\bf Analytical proof for intermediate mass scale $M_{\rm I}$:-}
The analytic formula for intermediate mass scale $M_{\rm I}$ due to one-loop, two-loop and gravitational corrections is read as,
\begin{eqnarray}
 {\large \ln}\left(\frac{M_I}{M_Z}\right)&&=\frac{B_{U} \pmb{D_{S}}-A_{U} \pmb{D_{W}}}{B_{U} A_{I}-B_{I} A_{U}} + \frac{A_{U} K_{\pmb{\Theta} \pmb{\Theta^\prime}}-B_{U} J_{\pmb{\Theta} \pmb{\Theta^\prime}}}{B_{U} A_{I}-B_{I} A_{U}} \nonumber \\
 &&
 -\frac{2\pi \Big(B_{U} \pmb{E_{1}}-A_{U} \pmb{E_{0}}\Big) \pmb{\alpha_G^{-1}}} {B_{U} A_{I}-B_{I} A_{U}}
 \label{rel:MI}
\end{eqnarray}
Now in order to prove the theorem for intermediate mass scale, we study the expression term by term.
\begin{itemize}
 \item One loop contribution to intermediate symmetry breaking scale read as,
 \begin{equation*}
 \left(\ln\frac{M_I}{M_Z}\right)_{1-loop}= \frac{B_{U} \pmb{D_{S}}-A_{U}\pmb{D_{W}}}{B_{U}A_{I}-B_{I}A_{U}}
\end{equation*}
Using the same logic as has been discussed in case of $\sin^2\theta_{W}$ , for $B_U=0$, the above expression reduces to
\begin{equation*}
 \left(\ln\frac{M_I}{M_Z}\right)_{1-loop}= \frac {\pmb{D_{W}}}{B_{I}}\, .
\end{equation*}
Here the parameter $\pmb{D_{W}}$ contains experimentally measured values like $\sin^2\theta_W$ and $\alpha_{\rm em}$. $B_{I}$ has been already shown to be independent of one-loop effects at $\mu > M_{\rm I}$, hence the proof of the theorem is natural to show the intermediate mass scale $M_I$ has vanishing contributions due to one-loop effect emerging from mass scale $\mu > M_{\rm I}$. 
\item Similarly the two loop effect on $M_I$ is exhibited in the 2nd term of eq.(\ref{rel:MI}) given as
\begin{equation*}
 \left(\ln\frac{M_I}{M_Z}\right)_{2-loop}= \frac{A_{U} K_{\pmb{\Theta} \pmb{\Theta^\prime}}-B_{U} J_{\pmb{\Theta} \pmb{\Theta^\prime}}}{B_{U} A_{I}-B_{I} A_{U}} 
\end{equation*}
Here for $B_U=0$ the expression reduces to 
\begin{equation}
 \left(\ln\frac{M_I}{M_Z}\right)_{2-loop}= 
   -\frac{K_{\pmb{\Theta} \pmb{\Theta^\prime}}}{B_{I}} \nonumber
\end{equation}
It has already been demonstrated that $K_{\pmb{\Theta} \pmb{\Theta^\prime}}$ and $B_{I}$ are independent of two-loop contributions emerging from mass scale $\mu > M_{\rm I}$.
\item The gravitational correction for $M_I$ as noted in the 3rd term of eq.(\ref{rel:MI}) is as follows,
\begin{equation*}
 \left(\ln\frac{M_I}{M_Z}\right)_{NRO}= -\frac{2\pi(B_{U}E_{1}-A_{U}E_{0})}{\alpha_G(B_{U}A_{I}-B_{I}A_{U})}
\end{equation*}
Here again for $B_U=0$ and $E_0=0$, we see that
	\vspace*{-0.3cm}
\begin{equation}
 \left(\ln\frac{M_I}{M_Z}\right)_{NRO}=0 \nonumber
 \end{equation}
 Thus the gravitational corrections has no effect on the intermediate mass scale. 
\end{itemize}
 
\noindent

The proof of the theorem is independent of the choice of particle content and thus, the proof is robust. It can be generalized to both supersymmetric as well as non-supersymmetric version of Grand Unified Theories that accommodates intermediate trinification symmetry $(G_{333D})$. 

We shall now show the stability of electroweak mixing angle $\sin^2\theta_W$ and intermediate mass scale $M_I$ in specific $E_6$ GUT models with or without supersymmetry by numerical analysis.
\noindent
\section{Predictions in SUSY $E_6$ GUT}
We consider first the supersymmetric version of $E_6$ GUT with intermediate trinification symmetry as,
\vspace*{-0.3cm}
{\small 
\begin{eqnarray}
	E_6\otimes SUSY&& \stackrel{M_U}{\longrightarrow} \mathbb{G}_{333D}\otimes SUSY \stackrel{M_I}{\longrightarrow} \mathbb{G}_{SM}\otimes SUSY\nonumber \\
	&& \hspace*{2cm}\stackrel{M_Z}{\longrightarrow} SU(3)_C\otimes U(1)_{Q}
\end{eqnarray}
}
\vspace*{-0.0cm}
At first stage, SUSY-$E_6$ is broken down to the trinification gauge symmetry $G_{333D}$ at unification scale $M_U$ by assigning non-zero vacuum expectation value (VEV) in the singlet direction $(1,1,1) \subset {650_H}$
which transforms evenly under D-parity. This singlet scalar preserves the symmetry between left and right handed Higgs field. The next stage of symmetry breaking from $G_{333D}$ to $G_{SM}$ is done by non-zero VEV of $(1, \overline{3},3)_{27}\oplus (1,3,\overline{3})_{\overline{27}}$. The last stage of symmetry breaking $G_{SM}$ to $G_{31}$ is done by the weak doublets $(1,2,1)_{27}\oplus (1,2,-1)_{\overline{27}} $ at $M_Z$ scale reproducing all known SM fermion masses. Here the supersymmetry is broken at the $M_Z$ scale.

   The usual MSSM particle spectrum are used for derivation of one-loop, two-loop beta coefficients  in the mass range $M_Z-M_{\rm I}$. The fermions contained in fundamental representation of $E_6$ i,e $27_F$ plus scalar sectors including $\Phi(1,\overline{3},3)+\overline{\Phi}(1,3,\overline{3})$ are considered in mass range $M_{\rm I}-M_U$. The derived values of one-loop beta coefficients are,
   $\pmb{b_{3C}}=-3, \pmb{b_{2L}}=1, \pmb{b_{Y}}=6.66$ in the mass range $M_Z-M_{\rm I}$ while $\pmb{b^\prime_{3C}}=0, \pmb{b^\prime_{3L}}=\pmb{b^\prime_{3R}}=3$ in the mass range $M_I-M_{\rm U}$. Similarly, the two-loop beta-coefficients derived for mass range $M_Z-M_{\rm I}$ and $M_{\rm I}-M_U$, respectively, are as follows
\begin{eqnarray}
\pmb{b_{ij}} =
 \begin{pmatrix}
  199/25 & 27/5 & 88/5 \\
  9/5    & 25   & 24    \\
  11/5   & 9    & 14
 \end{pmatrix}\, ,\quad  \pmb{b^\prime_{ij}}= \begin{pmatrix}
  82 & 40 & 24 \\
  40 & 82 & 24    \\
  24 & 24 & 48
 \end{pmatrix}\, . \nonumber
\end{eqnarray}

In the present case, the gravitational contribution will arise from the non-zero VEV of the Higgs $\phi(1,1,1)\subset 650_H\subset E_6$.
\vspace{-10pt}
\begin{eqnarray}
 & &\hspace*{-1.0cm} \langle \pmb{\Phi_{650}} \rangle=\frac{\langle\phi^0\rangle}{3\sqrt2} \textbf{Diag} \left\{\underbrace{2,\cdots,2}_{9},\underbrace{-1,\cdots,-1}_{9},\underbrace{-1,\cdots,-1}_{9}\right\} \nonumber
 \label{eq:650VEV}
\end{eqnarray}
From above equation, we have
$\pmb{\epsilon_{3C}} = 2 \pmb{\epsilon},
    \pmb{\epsilon_{3L}} =\pmb{\epsilon_{3R}}= - \pmb{\epsilon},
     \mbox{with}\,
  \pmb{\epsilon} = \frac{1}{3\sqrt{2}}\frac{\eta\, \langle \phi^0 \rangle}{M_G}$, which obviously conserves D-parity.\\
Now using the above values of one-loop beta coefficients, we estimated numerically the values of $M_I$, $M_U$, $\alpha_G^{-1}$ and $\sin^2\theta_W$ for different values of gravitational corrections in terms of $\epsilon$ (we skipped the two loop effects for simplicity) as presented in Table.\ref{tab:susyresults}.  
\begin{table}[h]
\centering
\vspace{-2pt}
\begin{tabular}{||c|c|c|c|c||}
\hline \hline
$\pmb{\epsilon}$ & $M_I$\,\mbox{(GeV)} & $M_U$\,\mbox{(GeV)} & $\alpha_{G}^{-1}$ & $\sin^2\theta_W$\\
\hline
$0$ & $1.691\times10^{16}$ & $5.044\times10^{16}$ & $24.0851$ & $0.2329$\\
\hline
$0.01$ & $1.691\times10^{16}$ & $2.225\times10^{17}$ & $23.6125$ & $0.2329$\\
\hline
$0.025$ & $1.691\times10^{16}$ & $1.853\times10^{18}$ & $22.9375$ & $0.2329$\\
\hline
$0.05$ & $1.691\times10^{16}$ & $4.906\times10^{19}$ & $21.3328$ & $0.2329$\\
\hline \hline
\end{tabular}
\caption{Numerical results for $\sin^2\theta_W$, $M_{\rm I}$, $M_U$ with different values of gravitational correction within SUSY-$E_6$ GUT with intermediate trinification symmetry.}
\label{tab:susyresults}
\end{table} 
\\
The above results show that the intermediate mass scale $M_I$ as well as $\sin^2\theta_W$ are not affected by the gravitational correction.
\noindent
\section{Predictions in NON-SUSY $E_6$ GUT}
The non-SUSY version of $E_6$ GUT with intermediate trinification symmetry is given by
\begin{eqnarray}
	E_6 \stackrel{M_U}{\longrightarrow} \mathbb{G}_{333D} 
	\stackrel{M_I}{\longrightarrow} \mathbb{G}_{SM}
	\stackrel{M_Z}{\longrightarrow} \mathbb{G}_{31}
	\end{eqnarray}
We have used $650_H$ as well as $27_H$ for spontaneous symmetry breaking 
while all the fermions are contained in fundamental representation $27_F$ of $E_6$ which accommodates all 15-SM component per generation. All the particle content of SM with Higgs boson have been used in the mass range $M_Z-M_I$ while all the fermions belonging to $27_F$ and the scalar $\Phi(1,\overline{3},3)$ are used in the mass range of $M_I-M_U$. 
The derived values of one-loop beta coefficients are,
   $\pmb{b_{3C}}=-7, \pmb{b_{2L}}=-19/6, \pmb{b_{Y}}=41/10$ in the mass range $M_Z-M_{\rm I}$ while $\pmb{b^\prime_{3C}}=-5, \pmb{b^\prime_{3L}}=\pmb{b^\prime_{3R}}=-9/2$ in the mass range $M_I-M_{\rm U}$. 
The two-loop beta-coefficients derived for mass range $M_Z-M_{\rm I}$ and $M_{\rm I}-M_U$, respectively, as follows
\begin{eqnarray}
\pmb{b_{ij}} =
 \begin{pmatrix}
  199/50 & 27/10 & 44/5 \\
  9/10    & 35/6   & 12    \\
  11/10   & 9/2    & -26
 \end{pmatrix}\,, \pmb{b^\prime_{ij}}= \begin{pmatrix}
  23 & 20 & 12 \\
  20 & 23 & 12    \\
  12 & 12 & 12
 \end{pmatrix}\,
\end{eqnarray}

\begin{table}[h]
\vspace*{-0.2cm}
\centering
\vspace{10pt}
\begin{tabular}{||c|c|c|c|c||}
\hline \hline
$\pmb{\epsilon}$ & $M_I$\,\mbox{(GeV)} & $M_U$\,\mbox{(GeV)} & $\alpha_{G}^{-1}$ & $\sin^2\theta_W$\\
\hline
$-0.0365$ & $1.18712\times10^{13}$ & $5.35202\times10^{16}$ & $47.0441$ & $0.2314$\\
\hline
$-0.034$ & $1.18712\times10^{13}$ & $5.18769\times10^{17}$ & $48.732$ & $0.2314$\\
\hline
$-0.0315$ & $1.18712\times10^{13}$ & $5.95424\times10^{18}$ & $50.5449$ & $0.2314$\\
\hline
$-0.030$ & $1.18712\times10^{13}$ & $2.81488\times10^{19}$ & $51.6989$ & $0.2314$\\
\hline \hline
\end{tabular}
\caption{Proof of theorem via numerical results within non-SUSY-$E_6$ GUT with intermediate trinification symmetry.}
\label{tab:nonsusyresults}
\end{table}
\vspace*{-0.3cm}
The numerically estimated values of $M_I$, $M_U$, $\sin^2\theta_W$ and $\alpha_G^{-1}$ for different values of $\epsilon$ have been presented in Table  \ref{tab:nonsusyresults}. We see that the intermediate mass scale $M_I$ as well as $\sin^2\theta_W$ are not affected by the gravitational correction. Here the unification mass scale $M_U$ and the GUT gauge coupling constant $\alpha_G$ changes with different values of $\epsilon$.   \\
In the present model the minimal particle content of $E_6$ is not enough for proper gauge coupling unification\cite{Stech:2003sb} at $\epsilon=0$, which comes out to be the scale beyond Planck energy. However, if we introduce $(1,8,8)\subset 650_H$, we achieve the unification at $10^{14}$ GeV. In order to achieve for a long proton lifetime the unifcation mass scale has to be enhanced with finite value of $\epsilon$ as well as with threshold effects\cite{Babu:2015bna}.\\


\noindent 
\section{Conclusion} 

In summary, we have proved a theorem on vanishing contributions to the electroweak mixing angle $\sin^2\theta_W$ and intermediate mass scale $M_{\rm I}$ (equivalent to the scale at which the spontaneous breaking of the trinification gauge symmetry $SU(3)_C\otimes SU(3)_L\otimes SU(3)_R\otimes D$ occurs), due to one-loop, two-loop and gravitational corrections emerging from higher mass scales ($\mu > M_{\rm I}$) in a class of grand unified theories which accommodates trinification symmetry invoked with spontaneous D-parity mechanism. We have established the robustness of the proof by considering supersymmetric as well as non-supersymmetric version of $E_6$ GUT for demonstration purpose. This proof can be generalized to show that there is no effect of GUT-threshold correction on these parameters arising from the mass scale $\mu > M_{\rm I}$. In order to prove this, the $SO(10)$ GUT has to be replaced by $E_6$ GUT and the intermediate Pati-Salam symmetry $\mathbb{G}_{224D}$ has to be replaced by trinification symmetry ($\mathbb{G}_{333D}$). The detailed proof is beyond the scope of this paper which is planned for a separate work. 

We, finally, conclude that the origin behind these vanishing contributions to $\sin^2\theta_W$ and intermediate mass scale $M_{\rm I}$ are primarily because of: 
(a) Grand Unified Theories like $E_6$ GUT that accommodates trinification symmetry as an intermediate breaking symmetry.
(b) Due to presence of discrete left-right symmetry (D-parity) and the implications of spontaneous D-parity breaking mechanism~\cite{Chang:1983fu,Chang:1984uy} thereby resulted simplified relations for the proof, 
 (c) Due to key matching condition between gauge couplings, 
 $  \alpha^{-1}_{Y}(M_I) = \frac{1}{5} \alpha^{-1}_{3L}(M_I) + \frac{4}{5}\alpha^{-1}_{3R}(M_I)\,$. It can also be explained in the GUT models like $SU(9)$ and $SO(18)$ with the trinification symmetry at the intermediate breaking symmetry~\cite{Hati:2017aez}.
 
 With reference to the cosmological issues, it is noteworthy to mention that the present model doesn't have the Domain-Wall problem as the intermediate mass scale $M_I$ is high\cite{Patra:1991dy} with  $10^{13}$ GeV for Non-Susy and $10^{16}$ GeV for SUSY cases(as shown in Table\ref{tab:susyresults}, \ref{tab:nonsusyresults})\cite{Kibble:1982dd}. As far as proton decay is concerned, the model with high $M_U$ allows a stable proton.

\vspace*{-0cm}
\noindent
\section*{Acknowledgments}
\vspace*{-0.0cm}
Chandini Dash is grateful to the Department 
of Science and Technology, Govt.of India for INSPIRE Fellowship/2015/IF150787 in support of her research work. She is also thankful to Dr.Sudhanwa Patra and 
IIT Bhilai for the kind hospitality where part of this work is completed.
\vspace*{-0.0cm}
\appendix
\section{Formalism for RGEs of gauge couplings in GUTs with trinification symmetry.}
\vspace*{-0.4cm}
We already proved the theorem showing the remarkable property of $E_6$ GUT or all possible grand unified theories with $\mathcal{G}_{333D}$ trinification intermediate symmetry on vanishing contributions to the electroweak mixing angle $\sin^2\theta_W$ and intermediate symmetry breaking scale $M_{I}$ due to one-loop, two-loop and gravitational corrections (even can be true for GUT-threshold corrections). The attempts have also been made to show that GUT threshold corrections arising out of super heavy masses or higher dimensional operators identically 
vanish on $\sin^2\theta_W$ or the $G_{224D}$ breaking scale within $SO(10)$ models with $\mathcal{G}_{224D}$ Pati-Salam intermediate symmetry~\cite{Pati:1974yy,Mohapatra:1974gc,Senjanovic:1975rk} and we rather carried out our analysis in $E_6$ GUT with $\mathcal{G}_{333D}$ intermediate trinification symmetry. We aim to derive all the necessary analytic formulas which have been used in the text for proof of the theorem. The simple symmetry breaking chain considered here with intermediate trinification symmetry is given by
\begin{eqnarray}
\mathbb{G}_{\rm GUT} &\stackrel{M_U}{\longrightarrow}& SU(3)_C\otimes SU(3)_L\otimes SU(3)_R\otimes D (G_{333D})\nonumber \\ 
	&\stackrel{M_I}{\longrightarrow}&SU(3)_C\otimes SU(2)_L\otimes U(1)_{Y} (G_{321})\nonumber\\
	&\stackrel{M_Z}{\longrightarrow}& SU(3)_C\otimes U(1)_{Q} (G_{31})
	\end{eqnarray}
It is also worth to mention here that $\pmb{b^\prime_{i}}\, (\mbox{i=3C,3L,3R})$ are one-loop beta coefficients from mass range $M_I-M_U$, $\pmb{b_{i}}\, (\mbox{i=3C,2L,1Y}$) are one-loop beta coefficients from mass range $M_Z-M_I$ while $\pmb{b^\prime_{ij}}\, (\mbox{i,j=3C,3L,3R})$, $\pmb{b_{ij}}\, (\mbox{i,j=3C,2L,1Y})$ are two-loop beta coefficients from mass range $[M_I-M_U]$ $[M_Z-M_I]$, respectively.  
We can deduce three key relations including one-loop, two-loop and gravitational corrections using equations (\ref{rge-alphainv}),(\ref{rge-alphainv-muMI}) and (\ref{rge-alphainv-MIMU})  as,
{\small 
\begin{eqnarray}
\alpha^{-1}_{3C} (M_Z)&&=\alpha^{-1}_{G} (1+\epsilon_{3C}) + \frac{\pmb b_{3C}}{2 \pi} {\large \ln}\left(\frac{M_I}{M_Z}\right) 
         \nonumber \\
&&  + \frac{\pmb b^\prime_{3C}}{2 \pi} {\large \ln} \left(\frac{M_U}{M_I}\right)          + \pmb{\Theta_{3C}}+\pmb{\Theta^{\prime}_{3C}}\,, 
   \label{rge:3C}  \\
\alpha^{-1}_{2L} (M_Z)&&=\alpha^{-1}_{G} (1+\epsilon_{3L}) + \frac{\pmb b_{2L}}{2 \pi} {\large \ln}\left(\frac{M_I}{M_Z}\right)
 \nonumber \\
&&+ \frac{\pmb b^\prime_{3L}}{2 \pi} {\large \ln} \left(\frac{M_U}{M_I}\right)+ \pmb{\Theta_{2L}}+\pmb{\Theta^{'}_{3L}}\, , 
   \label{rge:2L}       \\
\alpha^{-1}_{Y} (M_Z)&&=\frac{1}{5}\alpha^{-1}_{G} (1+\epsilon_{3L}) +\frac{4}{5}\alpha^{-1}_{G} (1+\epsilon_{3R})
 \nonumber \\
&&+ \frac{\pmb{b_Y}}{2 \pi} {\large \ln}\left(\frac{M_I}{M_Z}\right)+ \frac{\frac{1}{5} \pmb{b^\prime_{3L}} 
 + \frac{4}{5} \pmb{b^\prime_{3R}}}{2 \pi} 
            {\large \ln} \left(\frac{M_U}{M_I}\right)
 \nonumber \\
&&+\pmb{\Theta_{Y}} + \frac{\pmb{\Theta^{\prime}_{3L}}+4\pmb{\Theta^{\prime}_{3R}}}{5}\, , 
  \label{rge:Y} 
\end{eqnarray}
}

We used two more relations derived based on RGEs for the gauge coupling constants as,
\begin{eqnarray}
&&\alpha^{-1}_{em}\left(\sin^2\theta_W- \frac{3}{8}\right)= \frac{5}{8}\left(\alpha^{-1}_{2L}(M_Z)-\alpha^{-1}_{Y}(M_Z)\right)
\nonumber \\
&&8\left(\alpha^{-1}_s- \frac{3}{8}\alpha^{-1}_{em}\right)=8\alpha^{-1}_{3C}-3\alpha^{-1}_{2L}-5\alpha^{-1}_{Y}
\label{app:tworel}
\end{eqnarray}

Now using equations (\ref{rge-alphainv}),(\ref{rge-alphainv-muMI}) and (\ref{rge-alphainv-MIMU}) along with above relation (\ref{app:tworel}), we obtain the following analytic formulas for intermediate mass scale, unification scale and $\sin^2\theta_W$ as,
{\small
\begin{eqnarray}
 &&{\large \ln}\left(\frac{M_I}{M_Z}\right)=\frac{B_{U} \pmb{D_{S}}-A_{U} \pmb{D_{W}}}{B_{U} A_{I}-B_{I} A_{U}} + \frac{A_{U} K_{\pmb{\Theta} \pmb{\Theta^\prime}}-B_{U} J_{\pmb{\Theta} \pmb{\Theta^\prime}}}{B_{U} A_{I}-B_{I} A_{U}} \nonumber \\
 &&\hspace*{3cm}
 -\frac{2\pi \Big(B_{U} \pmb{E_{1}}-A_{U} \pmb{E_{0}}\Big) \pmb{\alpha_G^{-1}}} {B_{U} A_{I}-B_{I} A_{U}}
 \label{app:MI}  \\
 &&{\large \ln}\left(\frac{M_U}{M_Z}\right) = 
 \frac{A_{I} \pmb{D_{W}}-B_{I} \pmb{D_{S}}}{B_{U} A_{I}-B_{I} A_{U}} + \frac{B_{I} J_{\pmb{\Theta} \pmb{\Theta^\prime}}-A_{I} K_{\pmb{\Theta} \pmb{\Theta^\prime}}}{B_{U} A_{I}-B_{I} A_{U}} \nonumber \\
 &&\hspace*{3cm}
 -\frac{2\pi \Big(A_{I} \pmb{E_{0}}-B_{I} \pmb{E_{1}}\Big) \pmb{\alpha_G^{-1}}} {B_{U} A_{I}-B_{I} A_{U}}
 \label{app:MU} \\
 &&\sin^2\theta_W= \frac{1}{A_U}\Big[\frac{3}{8}A_U 
 +\left(\frac{\alpha_{\rm em}}{\alpha_s}-\frac{3}{8}\right)B_U 
\nonumber \\
&&\hspace*{2.1cm}+ \frac{\alpha_{\rm em}\left(A_{U}B_{I}-A_{I}B_{U}\right)}{16\pi}\ln\left(\frac{M_I}{M_Z}\right) \Big]
\nonumber \\
&&\hspace*{2.5cm}+ \frac{1}{A_U}\Big[\frac{\alpha_{\rm em}\left(A_{U} K_{\pmb{\Theta} \pmb{\Theta^\prime}} -B_{U} J_{\pmb{\Theta} \pmb{\Theta^\prime}}\right)}{16\pi} \Big]
\nonumber \\
&&\hspace*{2.5cm}+\frac{1}{A_U} \Big[\frac{\alpha_{\rm em} \left(A_{U} \pmb{E_{0}} - B_{U} \pmb{E_{1}}\right)}{8 \alpha_G}\Big]
\label{app:sinsqthetaw}
\end{eqnarray}
}

where the relevant parameters used in deriving all these key analytic formulas are as follows,
{\small
\begin{eqnarray}
&& \pmb{D_S}= 16\pi \left[\alpha_S^{-1}(M_Z)-\frac{3}{8}\alpha^{-1}_{\rm em}(M_Z) \right] \, \\
&& \pmb{D_W}= 16\pi \alpha^{-1}_{\rm em}(M_Z)\left[\sin^2\theta_W-\frac{3}{8} \right]  \\
%
&&J_{\pmb{\Theta} \pmb{\Theta^\prime}}= 2\pi \Bigg[\left (8\, \pmb{\Theta_{3C}}-3\, \pmb{\Theta_{2L}}-5\, \pmb{\Theta_{Y}} \right)
\nonumber \\
&&\hspace*{2.5cm}+\Big(8\, \pmb{\Theta^\prime_{3C}}-4\, \pmb{\Theta^\prime_{3L}}-4\, \pmb{\Theta^\prime_{3R}} \Big) \Bigg]\, \\
&&K_{\pmb{\Theta} \pmb{\Theta^\prime}}= 2 \pi \left[\left (5\,  \pmb{\Theta_{2L}}-5\, \pmb{\Theta_{Y}} \right)+\left (4\, \pmb{\Theta^\prime_{3L}}-4\,  \pmb{\Theta^\prime_{3R}} \right) \right] 
\label{app:Ktheta}\\
%
&&
\pmb{E_1}=\left (8\, \pmb{\epsilon_{3C}}- 4\, \pmb{\epsilon_{3L}} - 4 \, \pmb{\epsilon_{3R}} \right) 
\\
 &&\pmb{E_0}=\left (4\, \pmb{\epsilon_{3L}}-4 \, \pmb{\epsilon_{3R}} \right) \\
%
&&A_I= \Bigg[\Big(8\pmb{b_{3C}}-3\pmb{b_{2L}}-5\pmb{b_{Y}} \Big)
\nonumber \\
&&\hspace*{2.5cm}-\Big(8\pmb{b_{3C}^{'}}-4\pmb{b_{3L}^{'}}-4\pmb{b_{3R}^{'}} \Big) \Bigg] \\
&& A_U= \left (8\pmb{b_{3C}^{'}}-4\pmb{b_{3L}^{'}}-4\pmb{b_{3R}^{'}} \right)  \\
&&B_I=\left[\left (5\pmb{b_{2L}}-5\pmb{b_{Y}} \right)-\left (4\pmb{b_{3L}^{'}}-4\pmb{b_{3R}^{'}} \right) \right]  \, \\
&&B_U= \Big(4\, \pmb{ b^\prime_{3L} } - 4\, \pmb{ b^\prime_{3R} } \Big)
\end{eqnarray}
}
\noindent
These parameters are characterstics of one loop, two loop and gravitational corrections to the electroweak mixing angle $\sin^2\theta_W$ and the  intermediate mass scale while we focus on two of the parameters $B_I$ and $B_U$, in most of times, in the analysis.  With $\pmb{b^\prime_{3L}}=\pmb{b^\prime_{3R}}$, the factor $B_U$ vanishes exactly and $B_I$ is independent of one-loop effects operative at mass scale   $\mu > M_I$. Similarly, $K_{\pmb{\Theta} \pmb{\Theta^\prime}}$ can be shown to be independent of two-loop effects emerging from mass scale $\mu > M_I$. However, it is found that the unification scale and the GUT gauge coupling constant are fully dependent on these corrections.
\bibliographystyle{utphys.bst}
\bibliographystyle{utcaps_mod}
\bibliography{E6}
\end{document}